\documentclass[twocolumn,preprintnumbers,superscriptaddress,prb]{revtex4}
\usepackage{graphicx,bm,times,url}
\usepackage{amsmath}
\usepackage{amsfonts}
\usepackage{natbib}
\graphicspath{{./fig/}{./png/}}
\usepackage{color}

\newcommand{\xx}{\bm{x}}
\newcommand{\mm}{\mbox{\boldmath $m$} {}}
\newcommand{\Fig}[1]{Figure~\ref{#1}}
\newcommand{\kk}{\bm{k}}
\newcommand{\dd}{{\rm d} {}}
\newcommand{\FF}{\mbox{\boldmath $F$} {}}

\begin{document}

\title{Quantifying the Tangling of Trajectories Using the Topological Entropy}

\author{S. Candelaresi}
\author{D. I. Pontin}
\author{G. Hornig}
\affiliation{Division of Mathematics, University of Dundee, Dundee DD1 4HN, United Kingdom}

\date{\today}

\begin{abstract}
We present a simple method to efficiently compute a lower limit of the topological
entropy and its spatial distribution for two-dimensional mappings.
These mappings could represent either two-dimensional
time-periodic fluid fl+ows or three-dimensional magnetic fields, which are periodic in one direction.
This method is based on measuring the length of a material line in the flow.
Depending on the nature of the flow, the fluid can be mixed very efficiently
which causes the line to stretch.
Here we study a method that adaptively increases the resolution
at locations along the line where folds lead to high curvature.
This reduces the computational cost greatly which allows us to study
unprecedented parameter regimes.
We demonstrate how this efficient implementation allows the computation of the variation of the
{\it finite-time topological entropy} in the mapping.
This measure quantifies spatial variations of the
braiding efficiency, important in many practical applications.
\end{abstract}


\maketitle

\textbf{
Mixing of fluid flows occurs in oceanic flows and industrial production
(e.g.\ concrete mixing and chocolate mixing).
For two-dimensional flows we can quantify the degree of mixing using quantities
like the Lyapunov exponent or the topological entropy.
Here we use a lower bound for the topological entropy and present an efficient
and numerically accurate implementation which can be used to study flows, but
also tangling of magnetic field lines in e.g.\ tokamaks.
We show that our approach eliminates a large amount of unnecessary calculations
while still maintaining a high level of precision.
This approach is then used to compute a spatially dependent lower bound of the
topological entropy to identify regions of high mixing and contrast them with
areas of no mixing.
}

\section{Introduction}
We discuss the
concept of the \emph{topological entropy}\cite{Adler-Konheim-1695-114-309-TrAmMathSoc}
as a measure of mixing of fluid particles in a two-dimensional flow, or field line tangling
in a three-dimensional vector field.
The notion of topological entropy was developed in the context of dynamical systems,
and has been used in the field of fluid dynamics to understand
fluid mixing\cite{boyland2000, Budisic-Thiffeault-2015-25-8-Chaos,
Sattari-Chen-2016-26-3-Chaos}.
Such fluid mixing can be the stirring of a substance (solid or liquid)
with applications in engineering and production.
However, the need to quantify the mixing or tangling of trajectories
appears also in other fields.
For example, in plasma physics, particularly for magnetic fields in tokamaks
and spheromaks\cite{Petrisor-Misguich-2003-18-1085-ChaosSolFrac}, the tangling of magnetic field lines
is crucial for transport processes in the plasma, however, the
concept of topological entropy has rarely been used in this context (notable exceptions are
e.g.~\cite{Klapper}\cite{Childress}).
In this case the magnetic field lines in a tokamak can be interpreted as world lines of a
two-dimensional dynamical system.
To make that interpretation complete we need to assume that the field is static
and periodic in the direction of the field.

In the following section we will describe three different methods for
measuring the topological entropy of a two-dimensional flow (or three-dimensional periodic magnetic field).
Each of these is based on the interpretation of the entropy as the
exponential stretching rate of a material line in the flow.
In order to apply the first two methods, one requires knowledge of the
mapping at every point in the domain, while the final method deals with
the case in which only a finite number of trajectories of the flow or
field is known (i.e.~the mapping is known only as discrete points).

After discussing the methods for estimating the topological entropy, we use our new
method to explore the properties of two mappings that appear in the study of magnetic
field dynamics in plasmas.
Many laboratory and astrophysical plasmas are characterized by high magnetic
and fluid Reynolds numbers, and as a result exhibit turbulent dynamics.
This dynamics typically leads to a situation in
which the magnetic field is highly disordered, with field lines being
tangled or braided in a non-trivial manner.
We are motivated to study the topological entropy in the context of these fields because it is now
apparent that the nature of the field line tangling - i.e.\ the detailed
magnetic field topology - is crucial in determining the field and plasma
dynamics, in particular their relaxed state (e.g. \citep{Parker-1972-174-499-ApJ,
Berger-2001-55-3-LMathPhys,
pontin2011a,
Hansteen-Guerreiro-2015-811-106-ApJ,
Smiet-Candelaresi-2015-115-5-PRL}).
This is due to the conservation of various topological quantities.
If magnetic reconnection is allowed, such fields relax into a state of potentially different
field line topology (this topology being preserved in the absence of reconnection).
Reconnection changes the field line mapping, which therefore changes the stretching
behavior of the above mentioned material line.
Typically the field undergoes some simplification which leads to a reduction of the stretching rate.
In laboratory plasmas experimentalists are interested in the length of magnetic field lines
until they hit the device's divertor plates which helps in estimating the safety factor
\cite{Kroetz-Roberto-2012-54-4-PlasFus}.
Although the methods described here do not measure such lengths, they can be used to distinguish
chaotic regions generated by the magnetic field.

At the end {of this paper we
propose a new measure that allows quantification of the local (in time)
tangling of trajectories, important in many applications -- the {\it finite-time topological entropy}.
Finally, we consider the distribution of a passive scalar to demonstrate the efficiency
of the mixing which will decrease length scales exponentially, provided the topological
entropy is positive.

\section{Topological Entropy: Methods of Estimation} \label{sec: Methods}
\subsection{Interpretation in terms of material line stretching}
An exact value for the topological entropy of a field or flow can be
obtained only in special cases, for instance for some analytically
prescribed shearing motions \citep{boyland2000}, due to its complicated definition
which involves taking the limits of refinements of coverings of a topological space
\cite{Adler-Konheim-1695-114-309-TrAmMathSoc}.
Thus, for most practical purposes the entropy must be
estimated, and there exist different methods by which this estimation may be performed.
One such method makes use of symbolic dynamics \cite{Day-Frongillo-2008-7-4-SIAMAppDyn}
which computes a lower bound for the topological entropy.
In this paper we focus on methods that determine the topological entropy $h(f)$ of a mapping (homeomorphism)
\begin{equation}
f: M \rightarrow M,
\end{equation}
where $M$ is a compact subset of $\mathbb{R}^2$.
This mapping $f$ can be induced, for example, either by a two-dimensional time-periodic flow or by
the mapping of points between two planes connected by a static periodic magnetic field.

The topological entropy $h(f)$ is approximated by the exponent $h(f,\gamma)$ of the rate of
stretching of a material line $\gamma \subset M$ under the flow\citep{Newhouse-Pignataro-1993-72-5-JStatPhys},
\begin{equation}
h(f)\ge h(f, \gamma).
\end{equation}
Estimations of this stretching rate must be carried out by some computational method.
In particular, for chaotic flows this is a numerically demanding task.
In general the quality of the approximation of $h(f)$ by $h(f, \gamma)$ depends on the
choice of $\gamma$, and only a supremum over the set of all possible curves $\gamma$
would yield the exact value of $h(f)$ \cite{Newhouse-Pignataro-1993-72-5-JStatPhys}.
However, for the examples of chaotic flows considered below, the approximation, $h(f)\approx h(f,\gamma)$,
is very accurate since under a few iterations of the mapping $\gamma$ comes close to
every point of the chaotic domain.

In our approximation of the topological entropy we measure the length
of the line $\gamma$ after each application of the mapping $f$.
For a continuous, but time-periodic  map $f(t)$, $f(t+T)=f(t)$, we identify the number
$n$ with the number of periods $T$.
The quantity we wish to measure is
\begin{equation}
h(f,\gamma) =  \lim_{n \rightarrow \infty} \frac{1}{n} \ln^{+} \left| f^n(\gamma)\right|,
\end{equation}
where $\left| f^n(\gamma)\right|$ denotes the length of the curve $\gamma$ under $n$ iterations
of the mapping and $\ln^{+}(x) = \max(\ln x,0)$.
For a numerical evaluation of this expression we discretize the initial curve $\gamma$
using a high number of points $\xx_i$, $i \in [1, N]$, along the curve and measure the distances
$\delta^i_0$, $i \in [1, N-1]$ between those.
The corresponding distances under $n$ iterations are called $\delta^i_n$.
For the limit of $\delta^i_0 \rightarrow 0$ the numerical approximation becomes exact:
\begin{equation}
h(f,\gamma) = \lim_{n\rightarrow \infty} \left( \frac{1}{n}\lim_{\delta_0^i \rightarrow 0}
\ln^+ \left(\frac{\sum_i\delta_n^i}{\sum_i\delta_0^i}\right) \right).
\end{equation}
Here the total length of the line after $n$ iterations is $\sum_i\delta_n^i$
and its initial length is $\sum_i\delta_0^i$. Note that the denominator in the
logarithm is bounded and hence does not contribute to the value of the expression in the
limit $n \rightarrow \infty$.

Before going on below to describe algorithms for estimating the
stretching rate, we first note connections with the calculation
of the Lyapunov exponent.
While in our calculations we compute a lower limit for the topological
entropy by measuring the lengthening of a mapped line,
the Lyapunov exponent measures the exponential separation of two neighbouring
points $\xx_0$ and $\xx_1 = \xx_0 + \mu\mm$ under the mapping $f$,
with the real positive parameter $\mu$ and normalized directional vector $\mm$.
The point separation after the application of the mapping $n$ times is
\begin{equation}
\delta_n(\xx_0, \mu, \mm) = ||f^n(\xx_0 + \mu\mm) - f^n(\xx_0)||.
\end{equation}
The maximum Lyapunov exponent can then be written as
\begin{equation} \label{eq: lyapunov}
\lambda(\xx_0) = \max_{\mm} \lim_{n\rightarrow \infty} \left( \lim_{\mu \rightarrow 0} \frac{1}{n}
\ln \left(\frac{\delta_n(\xx_0, \mu, \mm)}{\delta_0(\xx_0, \mu, \mm)}\right) \right).
\end{equation}
Note that in the standard definition a continuous time $t$ is used instead of
a discrete $n$.

A relation between the so called {\em metric entropy} and the Lyapunov exponent was derived by
Pesin \cite{Pesin-1977-32-55-RusMatSurv, Young-2003-313-entropy}.
Using arguments from measure theory, it was shown that the metric entropy
is, in general, smaller or equal than the sum of the positive Lyapunov exponents
(Ruelle's inequality \citep{Ruelle-1978-9-83-BolSocBrasMat}).
For Riemannian measure on the manifold $M$ the equality between the two
quantities could be shown \citep{Pesin-1977-32-55-RusMatSurv}.
For a detailed discussion of the relations between the topological entropy, metric entropy,
and Lyapunov exponents, the reader is referred to the discussion of Young\cite{Young-2003-313-entropy}.

In practice we can only take the limit $\lim_{n\rightarrow \infty}$ for special cases.
In all other cases we therefore compute $h(f, \gamma)$ for a finite number of iterations $n$, finite
line length and finite initial point separations $\delta_0^i$.
Since the mappings that we consider herein are dense within the attractor,
the particular choice of initial
line $\gamma$ is not critical
because for chaotic mappings the mapped
curve $\gamma$ gets arbitrarily close to every point in the domain for sufficiently large
iteration $n$,
as long as the initial line intersects the attractor.
Since we can identify the iterations $n$ of our mapping with a discrete time of a two-dimensional
fluid flow, we call this quantity the finite time topological entropy (FTTE) and write it as
\begin{equation} \label{eq: entropy length}
h(f, \gamma, n_{\rm iter}) =
\frac{1}{n_{\rm iter}} \ln\left(\frac{\sum_i\delta^i_{n_{\rm iter}}}{\sum_i\delta_0^i}\right),
\end{equation}
with the integer $n_{\rm iter}$.

\subsection{Direct Method: Adaptive Algorithm}\label{sec:adaptive}
Here we introduce an efficient algorithm for estimating the
stretching rate of a material line under a given mapping.
For positive values of the topological entropy the
length of the material line $\gamma$ scales exponentially with
the number of iterations \cite{Newhouse-Pignataro-1993-72-5-JStatPhys}.
Hence, the challenge is to compute the exponential growth rate $h$ of the length $l$ of $\gamma$
under iterations of $f$,
which is a lower limit for the exact topological entropy
\citep{Newhouse-Pignataro-1993-72-5-JStatPhys}.
It is worth noting that
$h$ will in general depend on the choice of curve $\gamma$ -- to find the tightest
lower bound on the topological entropy of the flow we must find the maximal
stretching rate for all possible curves $\gamma$.

During our computational implementation of the above procedure, an equal
resolution everywhere along the curve at all iterations requires an
exponentially increasing number of points.
Moreover, one requires to analyze
the mapping for a significant number of iterations (typically $10$-$20$ for the
mappings considered herein, as described below) in order to obtain an
accurate estimate of the stretching exponent.

To combat this difficulty we introduce here an adaptive method that directly measures the exponential
line stretching rate using a reduced number of points for each iteration of the mapping
similar to \cite{Dritschel-1989-10-77-CompPhysRep, Mills-2009-35-2020-CompGeo}.
To understand how this optimization of the resolution of the line works,
note first that each iteration typically stretches and folds the material
line -- see e.g.\ \Fig{fig: Henon} and \Fig{fig: final line horizontal}.
At any iteration the curve will be composed primarily of long sections with relatively
low curvature, between which are line sections
 with very high curvature which need to be resolved in order
to determine the total length of the curve accurately.
For a fixed and initially equidistant distribution of points one
requires a very large number of points (trajectories) to resolve the thin folds.

In order to reduce the total computational cost we apply an adaptive
method which adds points where the curvature causes the angle between
two consecutive line segments to be less than $\cos(\alpha) = 0.99$,
where $\alpha$ is the angle.
Those two line segments are spanned by three points.
If the condition is fulfilled we add one point between the first and second points
and another between the second and the third points on the {\em initial} line and
apply the mapping on those new points.
This refinement is repeated as long as there are two consecutive line segments
for which $\cos(\alpha) < 0.99$ or until the refinement results only in
a relative change of $10^{-3}$ of the line length, which we call the relative length tolerance
$l_{\rm tol}$ parameter in our method.
We do not add points if the length of the {\rm mapped} line
segment is less than $10^{-5}$ in order to avoid issues related to the
machine precision.
That cut off number for the is the absolute tolerance $l_{\rm min}$.
While refinement in high curvature segments was already used by \cite{Dritschel-1989-10-77-CompPhysRep},
the cut off criteria for not adding points do not seem to appear in the literature.
Without them, the number of points can quickly rise well beyond what can be stored in
a computer.
This is somewhat mitigated by the ``contour surgery'' in \cite{Dritschel-1989-10-77-CompPhysRep}
which effectively creates short cuts in folds that are sufficiently close together.
However, such folds are an essential part of the advected material line and cannot
be discarded for the computation of the topological entropy.
In \cite{Dritschel-1989-10-77-CompPhysRep} the authors used a cubic spline interpolation
that is able to counter such potential inaccuracies.

\subsection{Stretching Rate Methods}
An alternative to the direct method discussed above for estimating the
line stretching rate is the set of algorithms proposed by Newhouse \& Pignataro
\cite{Newhouse-Pignataro-1993-72-5-JStatPhys}.
These methods estimate the stretching rate without explicitly fully
resolving the mapped line.
Conceptually, the idea is to estimate the growth of the line by
measuring the stretching of tangent vectors to the curve under the mapping.
After each iteration these tangent vectors are re-scaled before being mapped
forward again, and the overall expansion factor is given by the products of
the expansions during each iteration.
In this process any mapped vectors that contract (as could occur for example
across regions of high curvature) are neglected.
The algorithm follows the same conceptual procedure as a common method
of calculating Lyapunov exponents \cite{benettin1980}.

\subsection{Braid Entropy Estimations}
A further method of estimating the topological entropy is described by
Thiffeault \cite{Thiffeault-2010-20-1-Chaos}.
This method differs from the two previously presented in that it is
based on the assumption that one knows only a fixed number of trajectories
(field lines) originating at discrete starting points rather than having
access to the full flow information
(as is required if one wishes, for example, to add additional points in regions of high curvature).
The approach involves constructing a (mathematical) braid from these
trajectories, and then finding the minimum length of a material loop
that is constrained to wrap around the trajectories.
Using the method of Moussafir\cite{moussafir2006} one can encode such a
loop as a set of coordinates, which can then be used to evaluate the
minimum possible length of the material loop.
As such, the addition of more trajectories in the calculation will in
general lead to a longer loop and so the topological entropy of the
braid calculated in this way provides a lower bound to the topological
entropy of the full flow (corresponding to the limit of infinitely
many known trajectories, supposing that the optimal choice of $\gamma$ has been made).
For details of the algorithm and underlying theory the reader is
referred to Thiffeault \cite{Thiffeault-2010-20-1-Chaos}.
It is worth noting that in order to obtain an accurate estimation of
the entropy with a small number of trajectories using this method one
requires either to integrate the trajectories for a very long time or
to average over ensembles of trajectories
\cite{Thiffeault-2010-20-1-Chaos,Budisic-Thiffeault-2015-25-8-Chaos},
both of which are computationally expensive.
The method is implemented in the freely-available \verb+braidlab+ package \cite{braidlab}.

\subsection{Passive Scalar/Density}\label{sec: scalar}
If we consider mappings, like the ones discussed here, to be world lines in a 2+1 dimensional space
(two spatial and one temporal dimension) then we can consider the braiding
as mixing of fluid particles.
The degree of mixing is then reflected in the power spectrum of a passive scalar $c(\xx)$.
This scalar is constructed such that it initially has a constant gradient profile in $\mathbb{R}^2$.
Its Fourier transform is then simply
\begin{equation} \label{eq: passive scalar fourier}
\mathcal{F}\{c(\xx)\}(\kk) = \int_{V} c(\xx) e^{i\kk\cdot\xx}\ \dd^2 x,
\end{equation}
where the volume $V$ spans the two-dimensional plane.
From that we can compute the power spectrum by integrating over $k$-shells
of width $\delta k$
\begin{equation} \label{eq: passive scalar shells}
\hat{c}(k) = \int\limits_{k-\delta k/2}^{k+\delta k/2} \mathcal{F}\{c(\xx)\}(\kk)\ \dd^2 k.
\end{equation}

The power spectrum is easily calculated for every application of the braiding
once we know the transformed passive scalar distribution $c(\FF(\xx))$, where
$\FF(\xx)$ is the mapping.
Since we know $\FF(\xx)$ analytically we can also compute $c(\FF(\xx))$ analytically.

To test the mixing of this passive scalar under the mapping $\FF(\xx)$ we
impose an initial profile $c(\xx) = x + y$.
This corresponds to a simple gradient in both the $x$- and $y$-direction.

\section{Test cases}
\subsection{H\'enon Map}
We describe in this section some maps that we use to verify our new algorithm for estimating
the topological entropy. We then go on to implement our algorithm to explore the properties of these maps.
We first consider
the well studied H\'enon map \cite{Henon-1976-69-50-CommMathPhys}, given by
\begin{eqnarray}
x_{i+1} & = & y_{i} + 1 - ax_{i}^2 \nonumber \\
y_{i+1} & = & bx_{i},
\end{eqnarray}
with the parameters $a$ and $b$ and the iteration $i$.
Here we will use $a = 1.4$ and $b = 0.3$ for which this map exhibits
chaotic behavior.
(Note that the mapping is not area-preserving and therefore could not be
generated by an incompressible flow, except for $b=\pm 1$.)
Newhouse \& Pignataro \cite{Newhouse-Pignataro-1993-72-5-JStatPhys} used this mapping to study their
estimates for the topological entropy for different parameters $a$ and $b$.
For $a = 1.4$, $b = 0.3$ they estimated the topological entropy to have a value of $0.4640$.

\subsection{Blinking Vortex}
Building on previous work \citep{pontin2011a,pontin2016a}
on topology of magnetic braids
we use two very similar maps, which are generated by blinking vortices.
The first we call $E1$, which is defined by:
\begin{eqnarray} \label{eq: e3}
 & & \phi_1 = 2\sqrt{2\pi} \kappa \exp(-((x_i-1)^2+y_i^2)/2) \nonumber \\
 & & \tilde{x} = -y_i\sin(\phi_1) + (x_i-1)\cos(\phi_1) + 1 \nonumber \\
 & & \tilde{y} = (x_i-1)\sin(\phi_1) + y_i\cos(\phi_1) \nonumber \\
 & & \phi_2 = -2\sqrt{2\pi} \kappa \exp(-((\tilde{x}+1)^2+\tilde{y}^2)/2) \nonumber \\
 & & x_{i+1} = -\tilde{y}\sin(\phi_2) + (\tilde{x}+1)\cos(\phi_2) - 1 \nonumber \\
 & & y_{i+1} = (\tilde{x}+1)\sin(\phi_2) + \tilde{y}\cos(\phi_2),
\end{eqnarray}
where $x_i$ and $y_i$ are the coordinates, $\kappa$ is the twist parameter,
$\phi_1$ and $\phi_2$ the twisting angles from the left and right
twist, respectively, $\tilde{x}$ and $\tilde{y}$ the mapped coordinates after the
left twist only and $x_{i+1}$ and $y_{i+1}$ the mapped coordinates
after the second twist.
This formalism is chosen such that there is a twist around the
point $(-1, 0)$ with angle $\phi_1$ and then around $(1, 0)$
with angle $\phi_2$.

The second such blinking vortex mapping we call $S1$ and differs
only by the sign of the second angle, i.e.\
\begin{equation} \label{eq: s3}
\phi_2 = 2\sqrt{2\pi} \kappa \exp(-((\tilde{x}+1)^2+\tilde{y}^2)/2).
\end{equation}

\subsection{Standard Map}
The standard
map\cite{Greene-1979-20-6-JMathPhys, Morrison-2000-7-6-PhysPlasm}
(also called the Chirikov-–Taylor map)
is a mapping that can be derived from a magnetic field on a toroidal surface.
It is defined as
\begin{eqnarray} \label{eq: standard map}
\theta_{i+1} & = & (\theta_i - \kappa\sin{(2\pi \phi_i)}/(2\pi)) \mod 1 \nonumber \\
\phi_{i+1} & = & (\phi_i + \theta_{i+1}) \mod 1,
\end{eqnarray}
with $\theta_i, \phi_i \in [0, 1]$.
By taking ${\rm mod}\ 1$ we make sure that the mapping is periodic.
A Poincar\'e map for the standard map is shown in \Fig{fig: poincare standard},
where we can clearly see locations of periodic orbits.
There are also regions of the map (particularly around $\theta=0.5, \phi=0$)
where the Poincar\'e map appears highly disordered, and indeed
Greene\cite{Greene-1979-20-6-JMathPhys} demonstrated that for the parameter value
chosen these regions are stochastic.

\begin{figure}[t!]\begin{center}
\includegraphics[width=0.9\columnwidth]{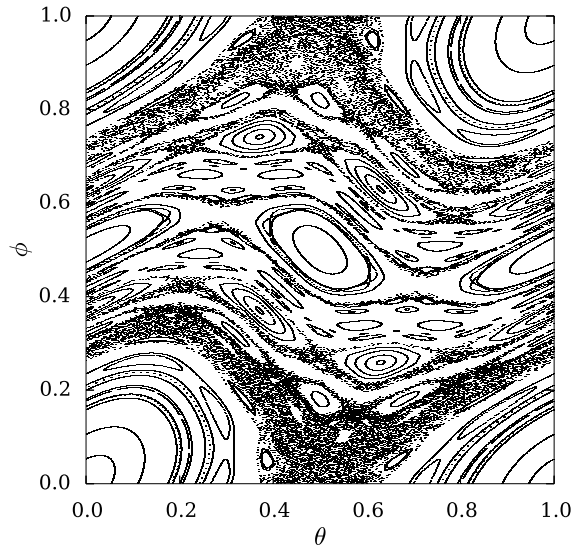}
\end{center}
\caption[]{
A Poincar\'e plot of the standard map \eqref{eq: standard map} for $\kappa =0.97$,
$500$ initial points and $1000$ iterations.
}
\label{fig: poincare standard}
\end{figure}

\section{Results}
\subsection{Benchmark of the Direct Method: H\'enon Map}
Since the H\'enon map was studied using previous
algorithms for estimating the topological entropy, we use it as a starting point to benchmark the new adaptive algorithm (of Section \ref{sec:adaptive}).
We start with the initial line close to the fixed point
at $(0.883, 0.883)$ extending from $(0.882, 0.883)$ to $(0.884, 0.883)$ (this choice being directed by the considerations
discussed in Newhouse \& Pignataro \cite{Newhouse-Pignataro-1993-72-5-JStatPhys}).
Applying our direct method  we obtain
an exponential increase in line length (see \Fig{fig: Henon}).
This is connected to the chaotic nature of the map, which leads to an exponential increase
of distance between neighboring points.
With increasing iteration step the number of points necessary to resolve
the entire mapped interval increases exponentially, and so does the computational
cost.
Here we are able to reach 25 iterations with limited computation time.
Similar to Newhouse \& Pignataro \cite{Newhouse-Pignataro-1993-72-5-JStatPhys} we fit the logarithm
of the mapped line length using linear fit and find a slope of ca.\ $0.46275$
(see figure \Fig{fig: Henon}, lower panel) which is close to the slope of
$0.4640$ found by Newhouse \& Pignataro \cite{Newhouse-Pignataro-1993-72-5-JStatPhys}.
This provides a confirmation of the accuracy of the new adaptive algorithm.

\begin{figure}[t!]\begin{center}
\includegraphics[width=0.9\columnwidth]{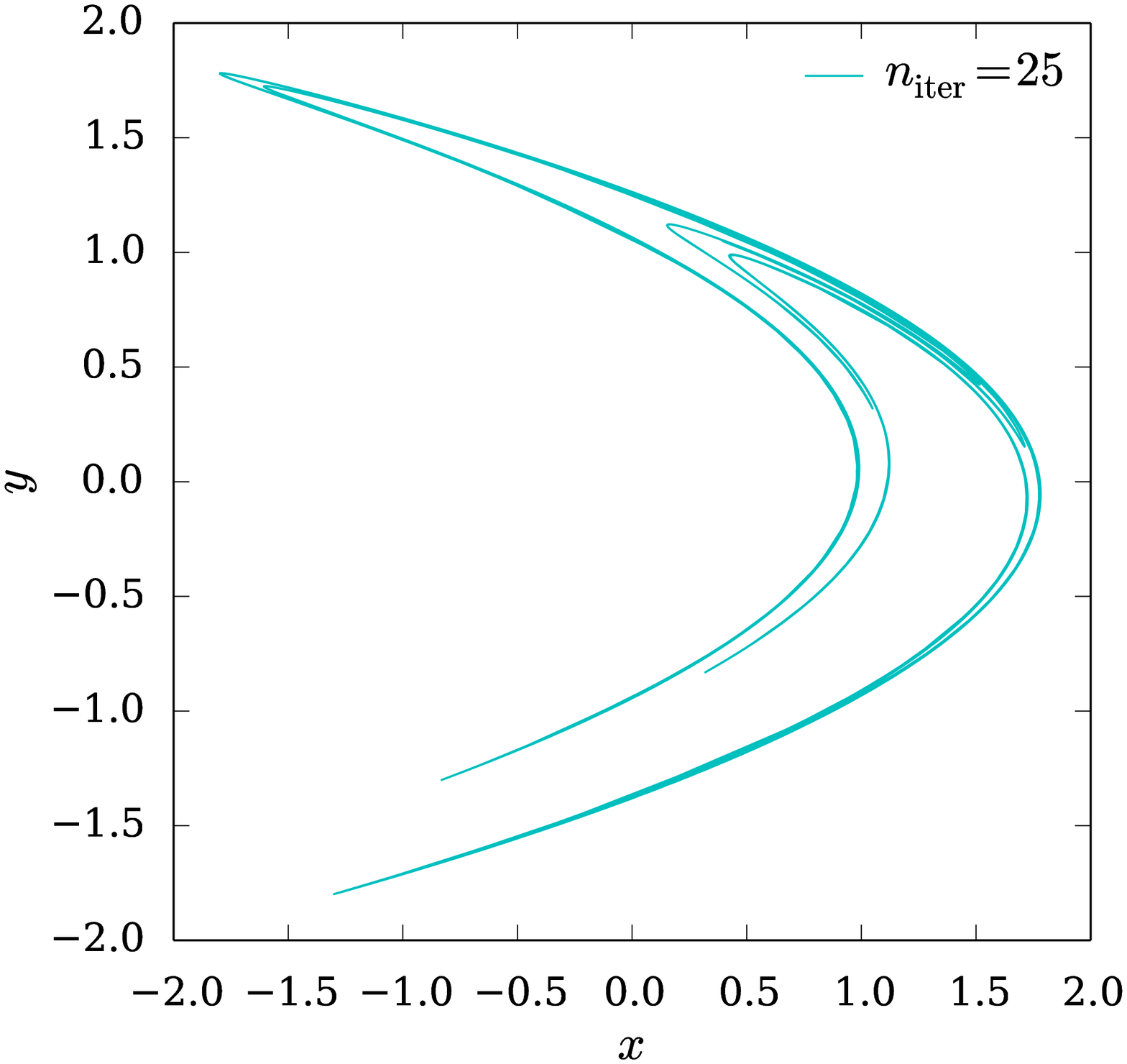} \\
\includegraphics[width=0.9\columnwidth]{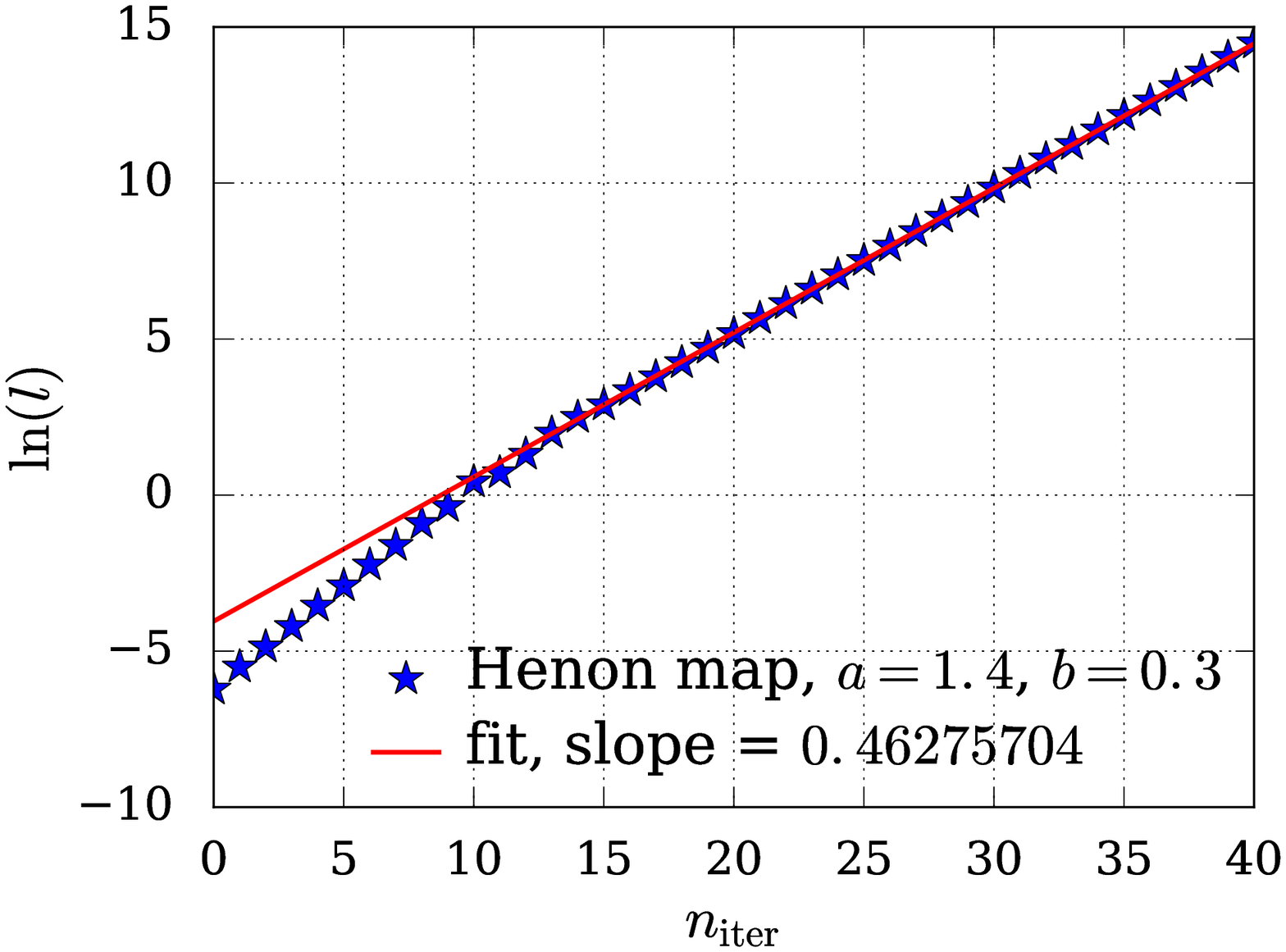}
\end{center}
\caption[]{
H\'enon map for $a = 1.4$ and $b = 0.3$ after 25 iterations (upper panel)
using the direct adaptive method.
Length of the line after $n$ iteration steps for the H\'enon map (lower panel)
with a linear fit for $\ln(l)$ with slope $0.46275$.
}
\label{fig: Henon}\end{figure}

\subsection{Direct Method: Blinking Vortex}
Moving now to consider the blinking vortex mapping in Eq.~\eqref{eq: e3},
we apply the direct method to an initial straight line $\gamma$ of length $4$, starting
at point $(-2, 0)$ and ending at $(2, 0)$.
In order to gain insights on the stability of our calculations we also perform
simulations with not just one initially horizontal line, but with $10$ different
straight lines of length 2 rotated about the origin.
The final length has a minor dependence on the initial line's orientation.
For each iteration we compute the mean line length of these $10$ initial conditions and the
corresponding standard deviation.

Depending on the parameter $\kappa$, the line gets highly tangled and shows
thin folds (\Fig{fig: final line horizontal}, upper panel).
Those folds are very frequent and need to be properly resolved in order
to measure the length of the line accurately.
With the adaptive method we achieve exactly this goal by increasing the
number of points at those segments (\Fig{fig: final line horizontal},
lower panel).
In order to achieve the same degree of accuracy with a fixed point
distribution, the number of points would need to be prohibitively
large.

\begin{figure}[t!]\begin{center}
\includegraphics[width=0.9\columnwidth]{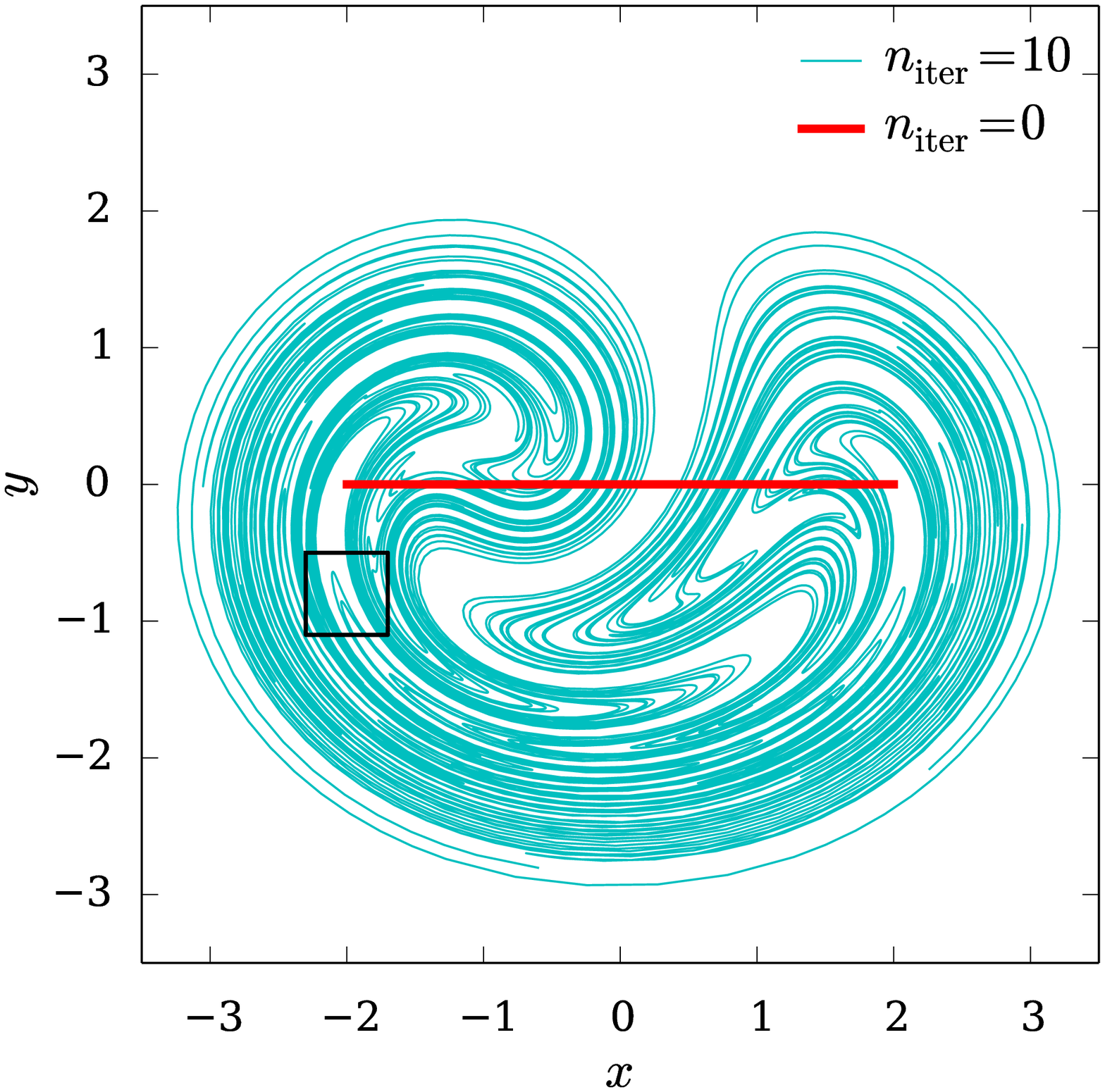} \\
\includegraphics[width=0.9\columnwidth]{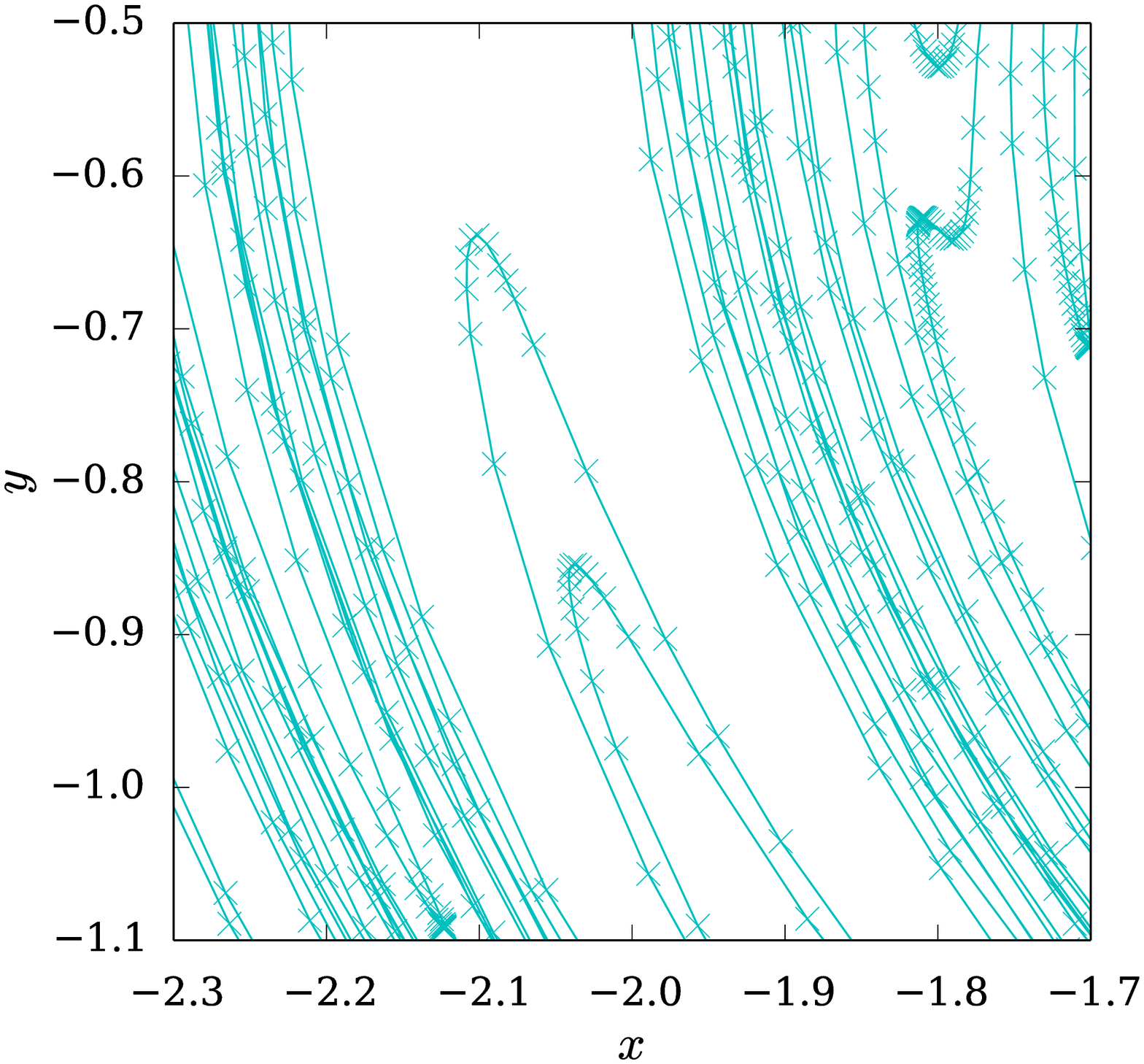}
\end{center}
\caption[]{
Blinking vortex $E1$ mapping of the initially horizontal line (thick red)
after $10$ iterations and $\kappa = 0.5$ (thin cyan).
The square denotes the zoom region for the lower panel where we also show
the distribution of points from the adaptive method.
It can be clearly seen that in segments where the curvature is high
the density of points is increased.
}
\label{fig: final line horizontal}\end{figure}

We plot the logarithm of the line length for different parameters $\kappa$
against the number of iterations $n_{\rm iter}$
(\Fig{fig: ln_l_vs_n_iter_star_e3}).
The length clearly follows an exponential law of type
\begin{equation} \label{eq: entropy fit}
l = a e^{h\,n_{\rm iter}},
\end{equation}
where the values of $a$ and $h$ depend on $\kappa$.
While the length of the stretched line is weakly sensitive on the orientation of the initial line
(by a few percent), this has little impact on the estimate of the entropy.
For the fits in \Fig{fig: ln_l_vs_n_iter_star_e3} we use the standard deviation together with
the mean values and obtain fits with tight confidence intervals between
$7.2903\times 10^{-4}$ for $h = 1.8585$ ($\kappa = 2.0$) and $4.4729\times 10^{-3}$ for $h = 2.3255$
($\kappa = 2.5$).
To compare with an equivalent non-adaptive implementation we compute $l$ for $\kappa = 2.5$
with the number of initially equally distributed points
corresponding to the final value of the adaptive method at each iteration.
While the adaptive method gives a clean exponential increase of $l$ with
$n_{\rm iter}$ the non-adaptive method starts flattening off at larger values of
$n_{\rm iter}$ (\Fig{fig: ln_l_vs_n_iter_star_e3}, triangles) which
is due to the presence of under-resolved line segments.

\begin{figure}[t!]\begin{center}
\includegraphics[width=0.9\columnwidth]{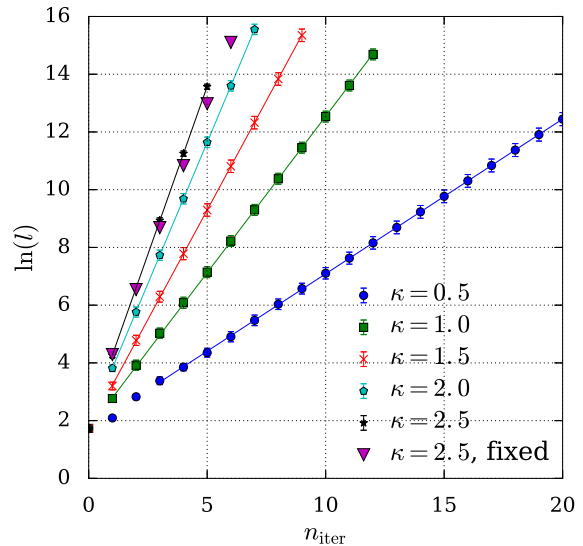}
\end{center}
\caption[]{
Mean logarithm of the mapped line length
versus the number of iterations for the $E1$ blinking vortex map for straight starting lines
crossing the origin and different values of $\kappa$ with the standard deviation.
We compare the results from the adaptive method with the fixed case (triangles).
The lines are least square fits for $l = a e^{h\,n_{\rm iter}}$, with fit
parameters $a$ and $h$.
}
\label{fig: ln_l_vs_n_iter_star_e3}\end{figure}

From the gradients in $\ln(l)$ (\Fig{fig: ln_l_vs_n_iter_star_e3}) we can
determine the lower limit, $h$, for the entropy for each value of $\kappa$, where $h$ is
the slope.
For the $E1$ mapping we obtain an almost linear increase of $h$ with $\kappa$
(\Fig{fig: slope_vs_k_e3_s3})
with absolute fitting standard deviations between $3.908\times 10^{-3}$ and $3.157\times 10^{-2}$.
For $\kappa = 1$ our results are consistently above the predicted lower bound of $0.9624$
by Boyland et al.\ \citep{boyland2000}.
For the $S1$ braid (see Eqs.~\ref{eq: e3},\ref{eq: s3}) we see a similar increase with a similar slope,
however we observe that the values are consistently below $E1$, consistent with
the theoretical considerations of Boyland\cite{boyland2000}.

To compare with the results of our algorithm we also perform calculations of
the finite-time braiding exponent (FTBE) for the same test cases using the \texttt{braidlab} package.
The results are very similar, as shown in Figure \ref{fig: slope_vs_k_e3_s3}
(with a standard deviation of $2\times 10^{-3}$ at $h = 1.0419$ for $\kappa = 1$).
For these calculations, we stack $60$ copies of the unit braid ($E1$ or $S1$), and use $10$ samples
of $N$ trajectories (from a set of $1200$) from this braid.
$N$ is increased from $500$ to $1000$ in steps of $50$, and the entropy estimated by an exponential
fit to the plot of mean FTBE versus $N$.
Even the step for the $S^1$ case at $\kappa = 1$ matches with the results from the
direct method.
Although the results are very similar, the computation times are very different.
To compare the computational efficiency of the two methods we measure the computation
time it takes to converge to a value for $h$ for the $E1$ braid with $\kappa = 1$.
For the direct method we use the computed lengths in \Fig{fig: ln_l_vs_n_iter_star_e3}
starting from $n_{\rm iter} = 3$ to compute $h$ and the linear fit.
By varying the total number of points used we obtain values of $h$ that approach
an asymptotic limit.
We then use the computation time for each $n_{\rm iter}$ and perform a fit of the form
$h(t) = h(1-\exp(-t/t_0))$, with the computation time $t$ and convergence time $t_0$.
Similarly, we vary the number of trajectories used in the \texttt{braidlab} calculations
from $200$ to $1000$, compute the value for the FTBE and measure the computation time.
We then use the same fitting function as for the direct method.
We run both methods for the $E1$ case with $\kappa = 1$ on 4 cores and observe
a convergence time of $380$s of computing for the \verb+braidlab+ package and
$0.83$s for the adaptive direct method.

\begin{figure}[t!]\begin{center}
\includegraphics[width=0.9\columnwidth]{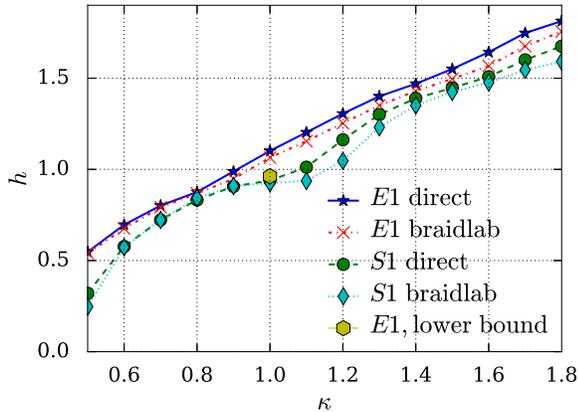}
\end{center}
\caption[]{
Estimate of the topological entropy for the $E1$ (solid blue line and asterisks) and $S1$ mapping
(dashed green line and circles) in dependence of the parameter $\kappa$
compared to the \texttt{bradilab} results for $E1$ (dash dotted red line and crosses)
and $S1$ (dotted cyan line and diamonds).
As a yellow hexagon we also plot the exact result for the lower bound for the $E3$ case
for $\kappa = 1$.
}
\label{fig: slope_vs_k_e3_s3}\end{figure}

In addition to the computation time, a further important consideration for measuring
an exponentially growing curve is the memory efficiency.
With increasing line length the number of points is expected to increase as well.
To test if that is the case for our adaptive method, we plot the
number of final points, after applying the mapping $n$
many times, with the line length for values of $\kappa$
between $0.1$ and $2.5$ and $n_{\rm iter}$ between $6$ (for higher $\kappa$)
and $150$ (for lower $\kappa$) (see \Fig{fig: npf_vs_l_fixed}, blue circles).
For comparison we also perform calculations with a fixed number of points that are equidistant.
We vary that number between $10^2$ and $10^8$ and measure the length of the stretched line for
this range of points.
As expected, as the number of points is increased, the length of the line increases
(as we obtain a higher resolution and avoid ``short cuts'').
We fit the logarithm of the line length in dependence of the number of (fixed and equidistant) points using
a function of the form $l(1-\exp(-\lambda n^{\rm p}_{\rm fixed}))$, where $n^{\rm p}_{\rm fixed}$
is the number of points.
From this we compute the number of necessary points (\Fig{fig: npf_vs_l_fixed}, red crosses) to
reach a value of $l$ $3\%$ away from its asymptotic value.
We clearly see that we obtain a different power law than linear which speaks against the efficiency of
such a fixed line approach.
Note that the value of $3\%$ can be easily changed.
However, the behaviour is the same (the power law becoming steeper for increasing desired accuracy).
The fact that with a fixed grid the number of points required grows faster than linearly with
the length of the line can be understood from the fact that, as the line length grows exponentially,
so the number of folds grows exponentially, and these folds  become (on average) tighter.
For the adaptive method,
the number of final points is clearly proportional to the line length
which means that the adaptive algorithm is efficient.
A further improvement would be to remove points where they are not
needed, i.e.\ along very straight line segments.
However, that would add computational complexity and we have
not implemented it here.

\begin{figure}[t!]\begin{center}
\includegraphics[width=0.9\columnwidth]{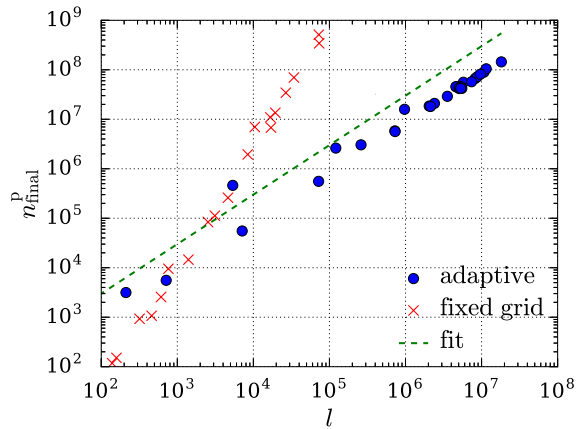}
\end{center}
\caption[]{
Number of final points for the adaptive method against the total
length of the line for a horizontal starting line and different
parameters $\kappa$ from $0.1$ to $2.5$ and iterations $n_{\rm iter}$
from $6$ to $150$ (blue circles)
together with similar calculations using a fixed number of points (red crosses)
and a linear fit (green dashed line).
}
\label{fig: npf_vs_l_fixed}\end{figure}

\section{Local measures of mixing efficiency}\label{sec: mixing}
So far we dealt solely with estimating the topological entropy, the notion
of which implicitly assumes a set of infinite trajectories (possibly within a periodic domain).
However, in many applications of interest this condition does not hold - that is
in practice we can only follow trajectories for a finite period of time.
Indeed, even when trajectories can be followed indefinitely, the finite-time
behavior may be more physically relevant.
In any fluid mixing process, one would like to determine the effectiveness of
the mixing over finite time.
If one considers the case of a magnetic field threading a plasma, that magnetic
field might not be embedded in a periodic domain.
Even if it exists in a periodic domain (for example a tokamak or spheromak)
the cases of interest are not stationary.
Therefore, any given magnetic field structure exists only for a finite period of
time, say $t\sim\tau$, and within this time period plasma particles can travel
some given finite number of times around the device.
As such, for a particle located at time $t=t_0$ on a given trajectory, one might
like to know how entangled that trajectory is with neighboring trajectories
for $t_0\leq t \leq t_0+\tau$.

The analysis of trajectories over finite time periods to determine their complexity
is typically done by calculating, for example, Finite-Time Lyapunov Exponents (FTLEs)
\citep{brunton2010}, a technique that has been used extensively in characterizing
unsteady fluid flows (e.g.\citep{Pierrehumbert1991}).
Such FTLEs measure only local deformation about a single trajectory.
Other examples of measures of local stretching or deformation in the mapping include
the finite time rotation number (FTRN)
\cite{Szezech-Caldas-2012-86-8-PRE, Szezech-Schelin-2013-377-452-PhysLetA},
the Mean Exponential Growth factor of Nearby Orbits (MEGNO) \cite{Cincotta-Simo-2000-147-205-AA}
and the Generalized Alignment Index (GALI) \cite{Skokos-Bountis-2007-231-30-PhysD}.

One approach that takes into account global (in space and time) changes is the
computation of the boundaries of Lagrangian coherent structures \cite{Haller-Yuan-2000-147-352-PhysD},
an approach that can be extended to arbitrary dimensions \cite{Lekien-Shadden-2007-48-6-JMatPhys}.
However, it involves the calculation of spatial derivatives of the flow which makes it
computationally less practical.

In additional to the local measures of stretching mentioned above,
a topological equivalent (measuring complexity on finite scales),
the so-called Finite-Time Braiding Exponent (FTBE),
was recently introduced by Budi\v{s}i{\' c} \& Thiffeault \cite{Budisic-Thiffeault-2015-25-8-Chaos}
(see also \cite{Thiffeault-2010-20-1-Chaos}).
A similar notion of finite-time entropy was introduced
\cite{Froyland2012} in the context of the {\it differential entropy},
to measure finite-time stretching.
Unlike the FTLE this measures (finite-time) non-linear stretching over a finite
$\epsilon$-neighbourhood of phase space -- though it reduces to the FTLE in the
limit $\epsilon\to 0$.
While this is similar in spirit to the FTTE that we discuss here, there are fundamental
mathematical differences between the two quantities; the finite-time entropy of
Froyland et al.\cite{Froyland2012}, being based on the differential entropy,
is more directly analogous to the metric entropy rather than the topological entropy.
For discussion on how the differential entropy relates to the metric entropy and
topological entropy see the review by Lesne\citep{lesne2014}.

\subsection{FTTE Distribution}
The topological entropy characterizes the tangling of chaotic trajectories by a single number.
If we wish to understand finite-time behavior of trajectories (or indeed if we only have
finite-time trajectory information), then we would naturally like to characterize the
trajectories locally in space as well as time (since we no longer have infinite trajectories
that fill (a portion of) the phase plane).
In this case, we propose that one may obtain useful information about behavior of the
system at hand  by analyzing the distribution of tangling within the domain.
One would like, for example, to take a grid of initial trajectories and evaluate
the local (in space and time) tangling per trajectory.
Within the framework presented in Section \ref{sec: Methods}, this can be achieved
by evaluating the FTTE over different curves $\gamma$ that cover $M$.
In particular, this distribution can tell us whether the domain is covered by multiple chaotic regions,
and whether these are separated or mixed with regions of non-chaotic trajectories.
This information is not given by the quantity $h$ which measures the maximum value of the topological
entropy on a given curve $\gamma \subset M$.
In order to gain additional information about the structure of the mapping we need to measure $h$ for curves
$\gamma$ which are embedded in distinct chaotic regions.

We first consider the blinking vortex mapping, in which the central portion of the $xy$-plane is
chaotic (for sufficiently large $\kappa$), but we know that at large distances from
the origin trajectories are not chaotic.
From \Fig{fig: final line horizontal} we see that the initial straight line is stretched out
over a significant portion of the central part of the plane, although there are also clearly
defined ``empty'' regions that the mapping does not reach.
If those areas are invariant for each iteration we should be able to measure
a different value of the FTTE in them from the surrounding regions where
the stretched and folded line densely fills the space.
In fact, we might measure vanishing entropy for some parts of the domain.
In order to measure the dependence of the FTTE on the coordinates
$x$ and $y$ we select an array of initial curves $\gamma$ given by small circles located on a rectangular grid.
These are then mapped forward as usual, and we monitor the
growth of the length of the corresponding curves with the methods described above.

For our experiment we seed $111\times 111$ circles within $x = [-4, 4]$ and $y = [-4, 4]$.
Their radius is chosen such that neighboring circles are sufficiently far apart
with radius $r = 0.8\times 8/(2\times 110)$, where $8$ is the size of the domain.
In order to increase the resolution we also perform computations with circles
at $x = [-4, 4]$ and $y = 0$ with $1001$ circles with radius
$r =  0.8\times 8/(2\times 1000)$.
We perform $20$ iterations of the blinking vortex map \eqref{eq: e3}
with $\kappa = 0.5$ and estimate the
FTTE using the fitting function for the line length as given in Eq.~\eqref{eq: entropy fit}.
Since the data is rather fluctuating below $n = 6$, we discard
points below this value when we perform the fit.

We find a clear distribution for $h$ with areas of $h \approx 0.6$
and areas with $h \approx 0$ (\Fig{fig: circles}).
It appears that the domain for which $h = 0.6$ has a fractal like structure and
resembles something similar to a Cantor set (compare the middle and lower frames in \Fig{fig: circles}).
The same is true for the domain for which $h = 0$.
Increasing the number of iterations improves the fit, and hence the calculation
of the entropy.
This also leads to a sharper divide with the range at $0 < h < 0.6$ decreasingly
populated.
Similarly with fractals, or the Cantor set, we also observe self-similarity when we
shrink the domain boundaries.

\begin{figure}[t!]\begin{center}
\includegraphics[width=0.9\columnwidth]{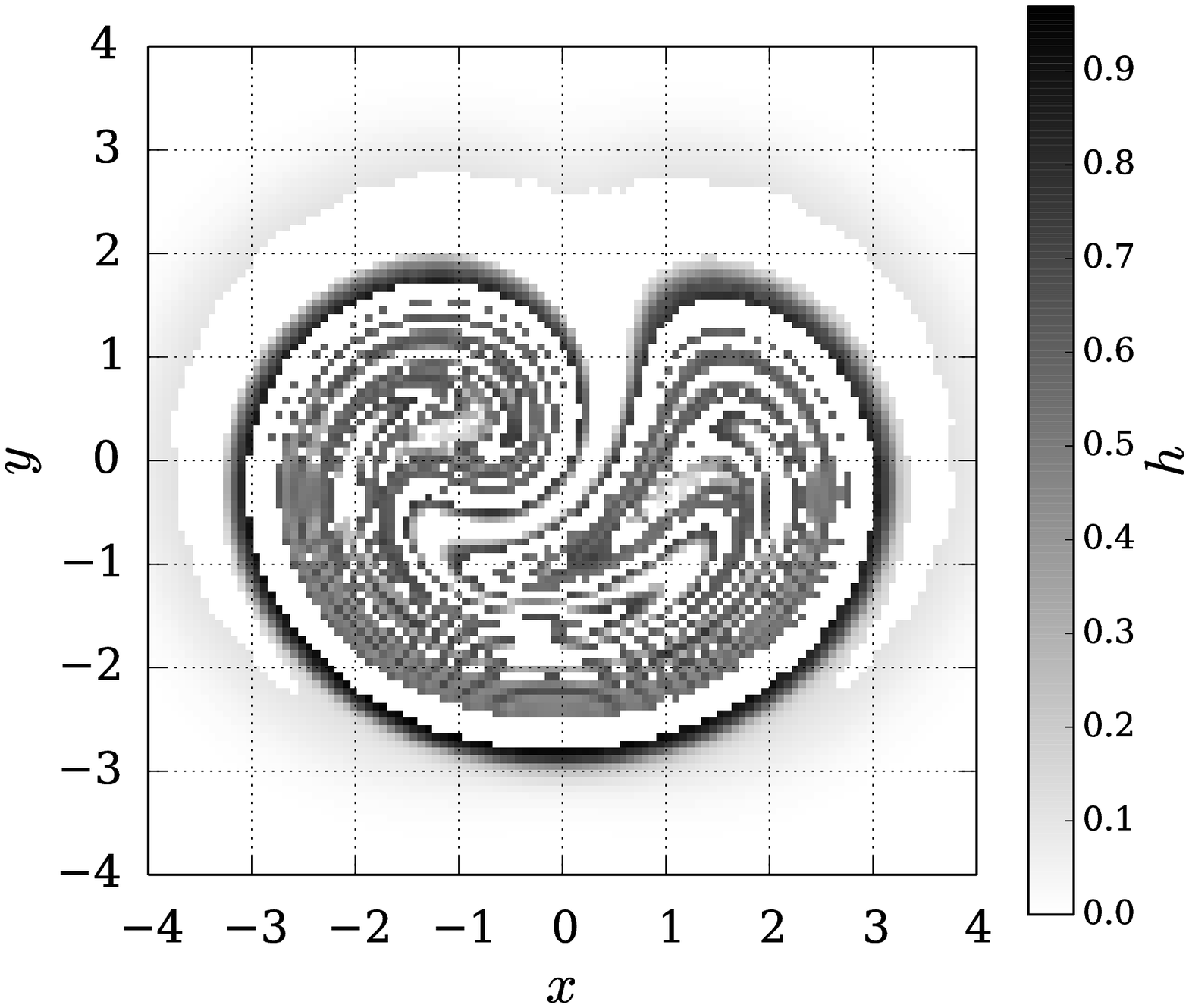} \\
\includegraphics[width=0.9\columnwidth]{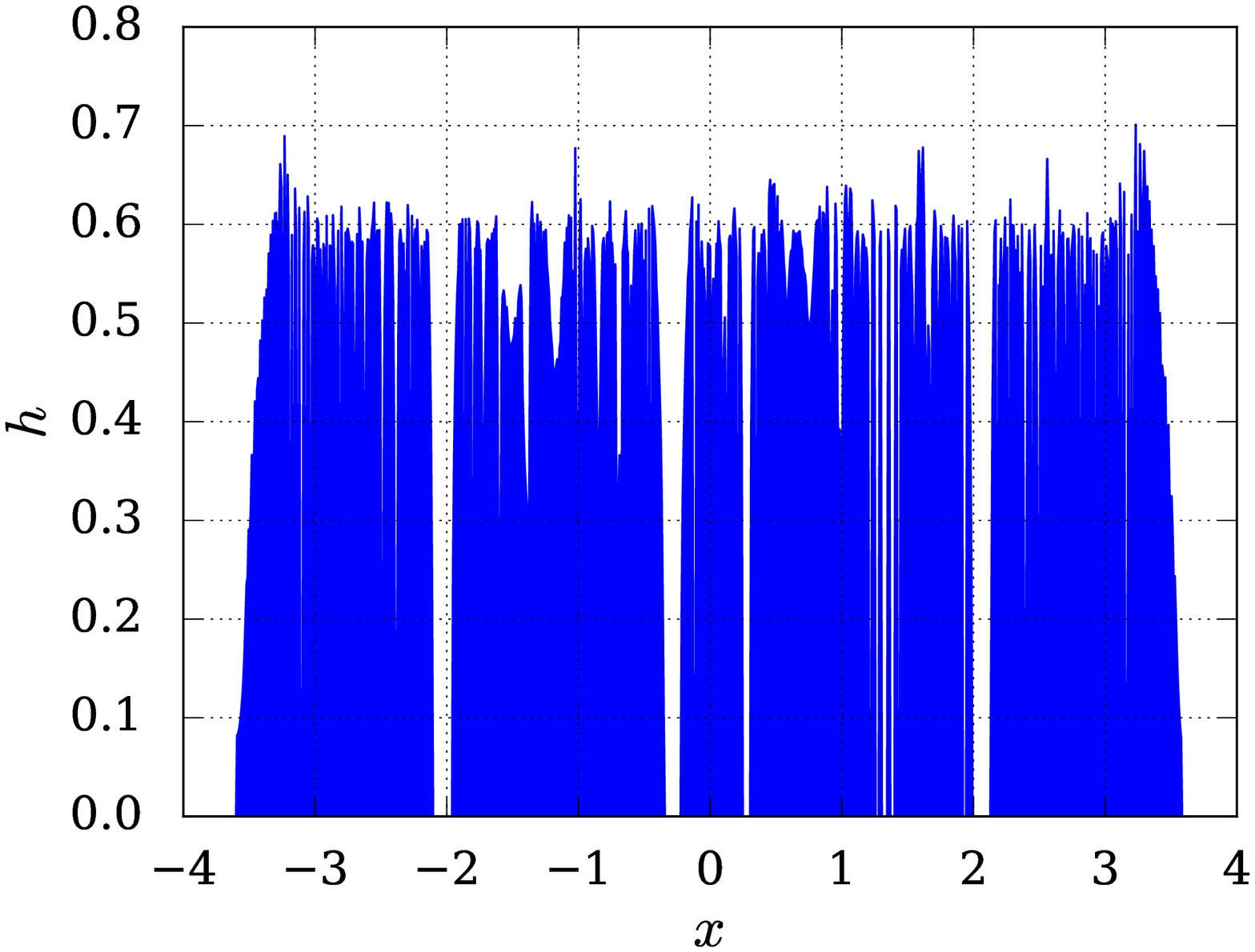}
\includegraphics[width=0.9\columnwidth]{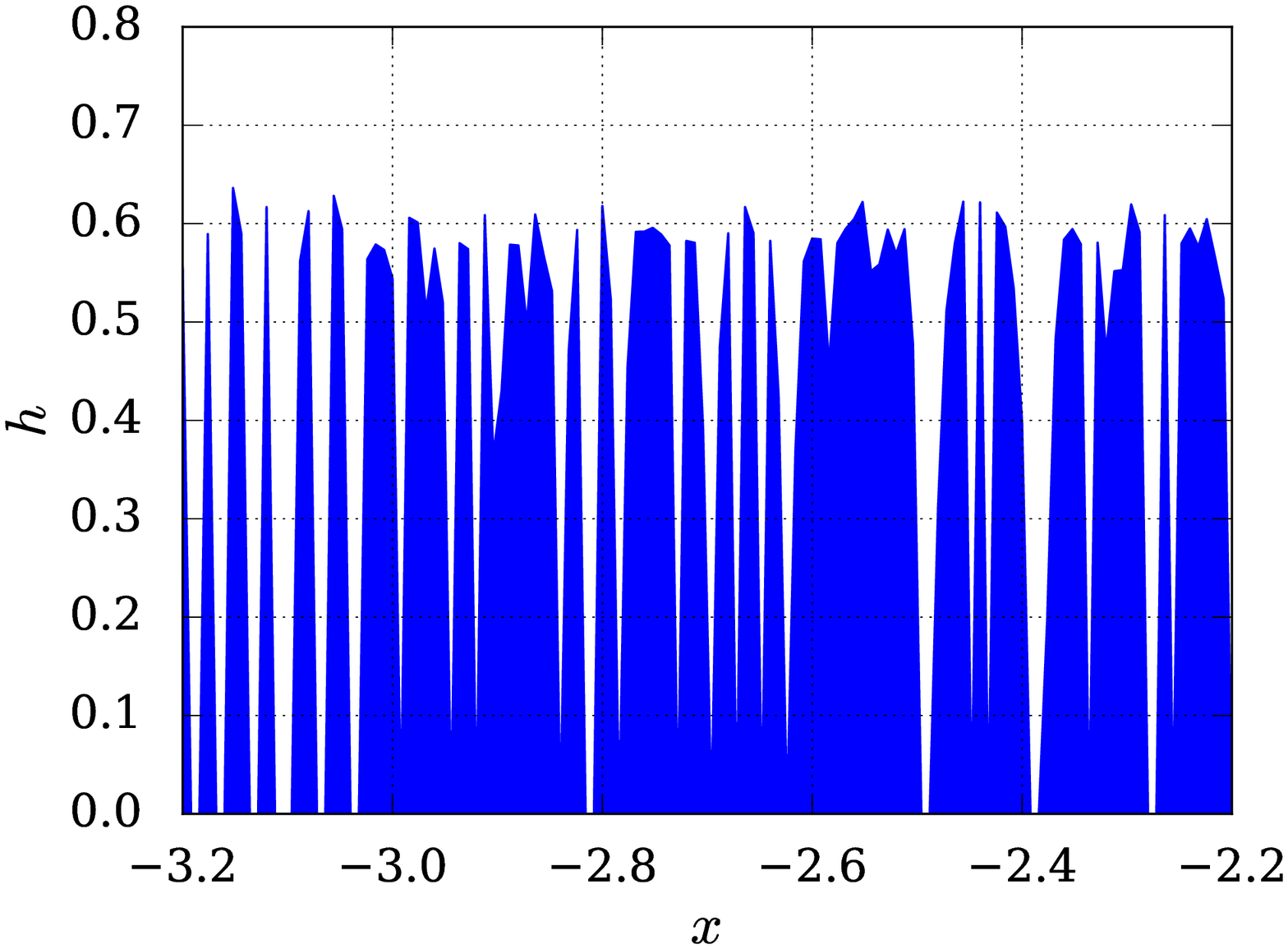}
\end{center}
\caption[]{
FTTE distribution for the blinking vortex mapping $E1$ with
$\kappa = 0.5$ and with initial
distribution of $111\times 111$ circles (upper panel), $1001$ circles at
$y = 0$ (central panel) and a zoom in of the region $-3.2 \le x \le -2.2$ (lower panel).
}
\label{fig: circles}\end{figure}

We also estimate the entropy distribution for the standard map
\eqref{eq: standard map} using the method of the initial circle distribution.
Intuitively we expect vanishing entropy at the periodic orbits, and non-zero
at chaotic regions.
It is not immediately clear, however, if there is any characteristic value for
the entropy.
From the FTTE distribution we can confirm our intuition, and find vanishing
entropy at periodic orbits (\Fig{fig: circles standard}).

\begin{figure}[t!]\begin{center}
\includegraphics[width=0.9\columnwidth]{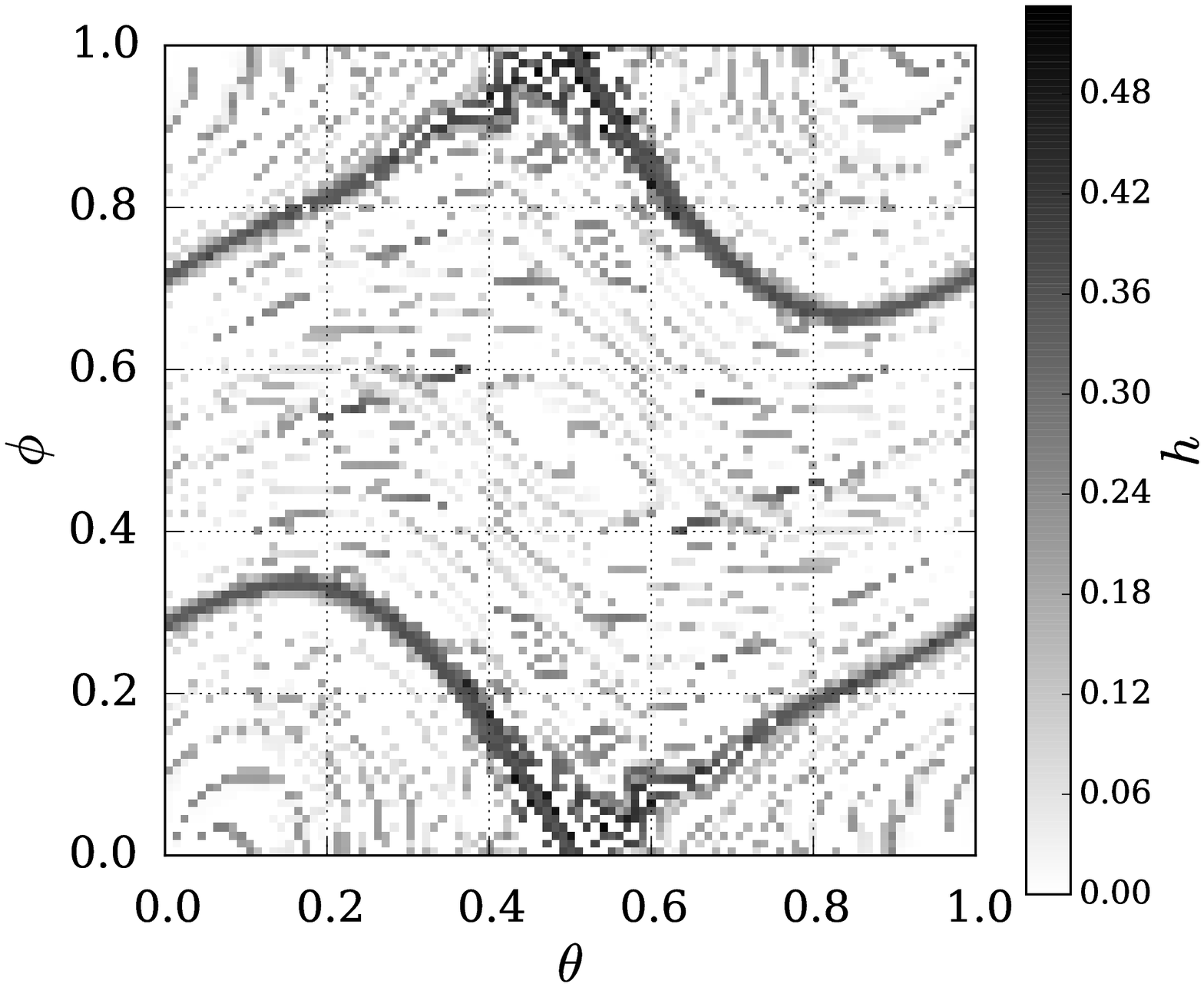} \\
\includegraphics[width=0.9\columnwidth]{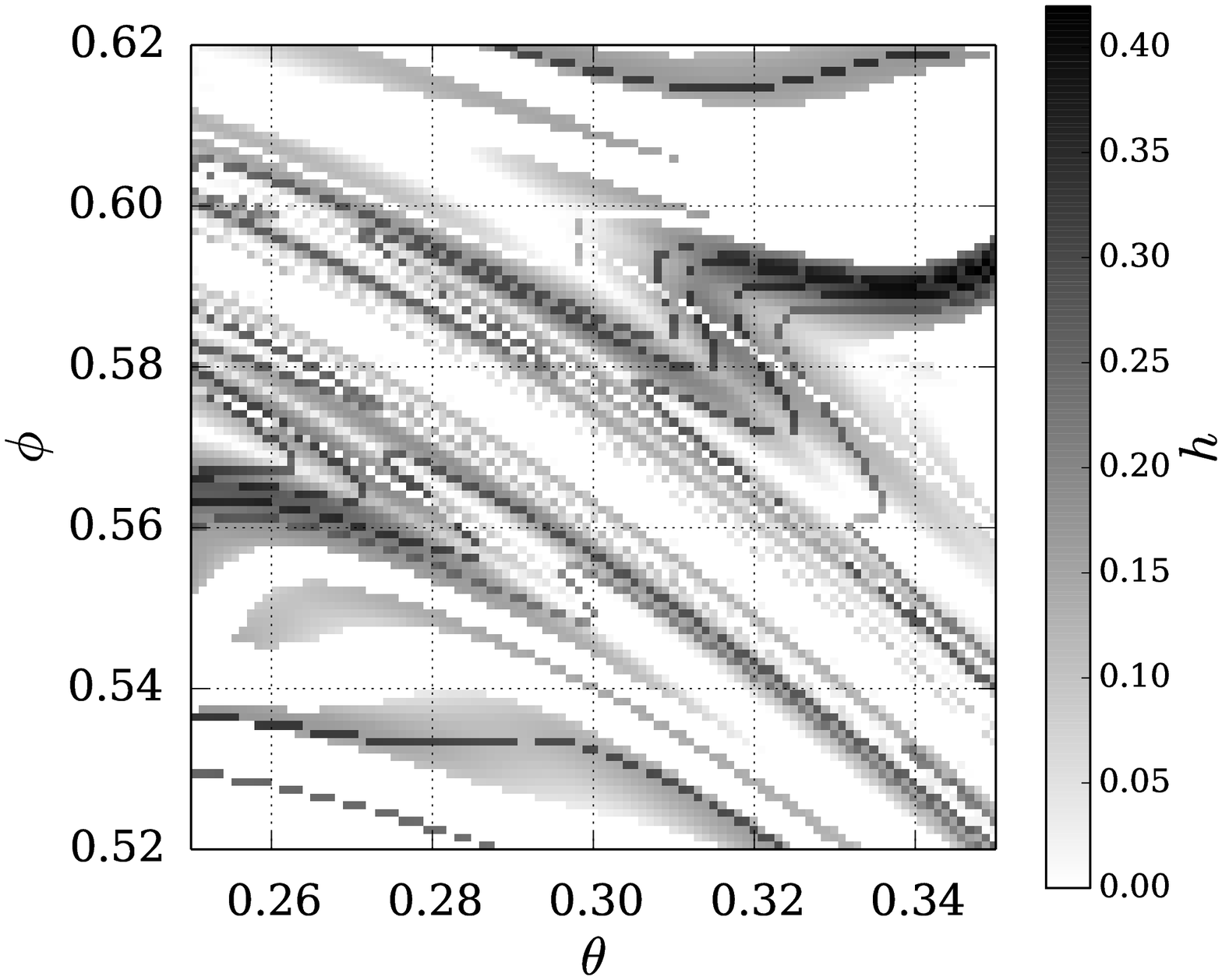}
\end{center}
\caption[]{
FTTE distribution for the standard map \eqref{eq: standard map}
and $\kappa = 0.97$ of $101\times 101$ circles (upper panel).
On the lower panel we show a zoom of the region $0.25 \le \theta \le 0.35$ and
$0.52 \le \phi \le 0.62$.
}
\label{fig: circles standard}\end{figure}

From the Lemma by Adler et al.\ \cite{Adler-Konheim-1695-114-309-TrAmMathSoc} we know that
for $\lim_{n_{\rm iter} \rightarrow \infty}$ the topological entropy of the mapping
is given as the maximum in the distribution.
This is exactly what we see when we compare \Fig{fig: slope_vs_k_e3_s3}
(for $\kappa = 0.5$) with \Fig{fig: circles}.
Intuitively, this is easily understood as the dominance of the largest growth rate.

For a magnetic field in a toroidal plasma the existence of distinct chaotic regions has far reaching implications.
Their boundaries determine regions which are not crossed by charged particles.
It also implies that, given the connection of $h$ and the finite time Lyapunov exponent,
magnetic field line separation behaves differently for different regions.
Such distinct regions were already computed in the past using the finite time rotation number
\cite{Szezech-Caldas-2012-86-8-PRE, Szezech-Schelin-2013-377-452-PhysLetA}.
They showed that it can be used to clearly distinguish disconnected regions that, in general,
have different values of the FTTE.

\subsection{Passive Scalar/Density}
To further probe the structure of the blinking vortex map we consider the
evolution of a passive scalar under successive iterations of the mapping as
discussed in Section \ref{sec: scalar}.
We first show that for the blinking vortex motion, the passive scalar
evolves identical to the fluid density.
If we replaced the passive scalar $c(\xx)$ by a density $\rho(\xx)$ the result
in equation \eqref{eq: passive scalar fourier} would be the same,
because our mapping $\FF(\xx)$ is volume preserving.
This can be easily shown by using the pull-back on the volume 2-form
$\dd x \wedge \dd y$:
\begin{eqnarray} \label{eq: cc two-form}
\FF^{*} (\dd x \wedge \dd y) & = &
\left( \frac{\partial F^1}{\partial x}\frac{\partial F^2}{\partial y}-
 \frac{\partial F^1}{\partial y}\frac{\partial F^2}{\partial x} \right)
 \dd x \wedge \dd y
 \nonumber \\
 & = & \dd x \wedge \dd y,
\end{eqnarray}
where $x$ and $y$ are the coordinates of the initial points.
Hence, the mapping $\FF(\xx)$ for the blinking vortex motion describes an incompressible flow.

Successive iterations strongly mix the initial distribution for the passive
scalar $c(\xx)$ (\Fig{fig: passive scalar}, upper panel).
The mixing is, however, not homogeneous.
We can identify regions with weak mixing and regions with strong mixing.
This is reflected in the power spectrum
(\Fig{fig: passive scalar}, bottom panel) where we clearly see
a noisy, but flat spectrum arising, which implies the non-existence of a characteristic
length scale.
The presence of a characteristic length scale would have shown up in
the power spectra.
Such a scale should have then decreased with successive applications
of the mapping.
Its absence shows that the stirring happens on all scales.
This is consistent with the identification of a fractal-like structure
to the FTTE distribution in the previous section.

It is remarkable that the shortest length scale with some measurable power changes
rapidly with the number of iterations.
In fact, we can see an exponential decrease in the smallest length scale,
as illustrated by the exponential function (\Fig{fig: passive scalar}, lower panel).
For a mapping with positive topological entropy, as in this case, we expected
such a behavior, since any initial line is being stretched exponentially
with the consequence that points on the line move exponentially close
to other points on the same line leading to thin structures.

\begin{figure}[t!]\begin{center}
\includegraphics[width=0.9\columnwidth]{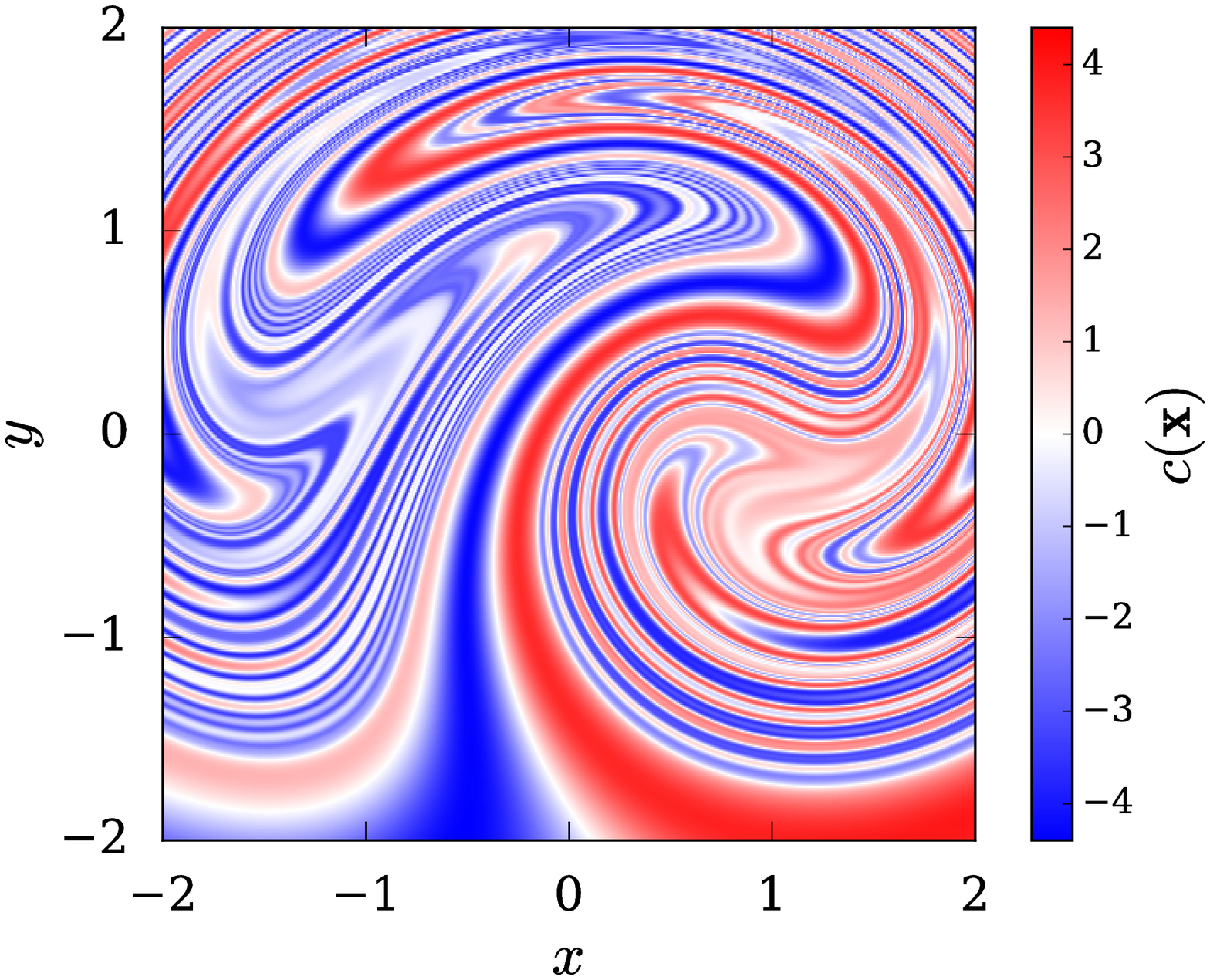} \\
\includegraphics[width=0.9\columnwidth]{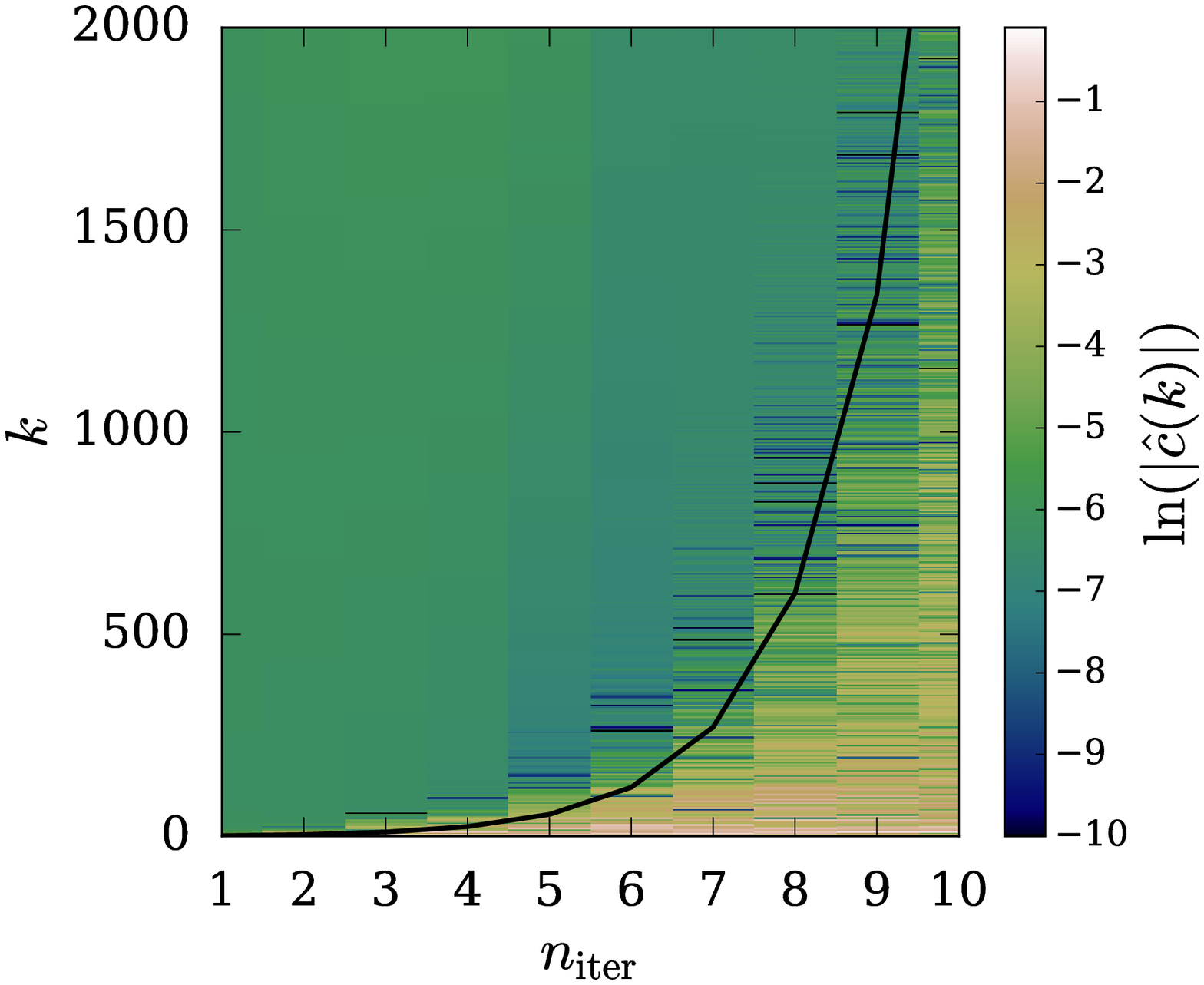}
\end{center}
\caption[]{
Passive scalar distribution after applying 7 iterations for the blinking
vortex mapping \eqref{eq: e3} with $\kappa = 0.5$ (upper panel).
Power spectrum for the passive scalar in dependence of the number
of iterations (bottom panel) together with an exponential curve (black line).
}
\label{fig: passive scalar}\end{figure}

\section{Conclusions}
Here we introduced a direct method to reliably and efficiently calculate
the lower limit of the topological entropy for a mapping $\FF: \mathbb{R}^2 \rightarrow
\mathbb{R}^2$.
Such mappings can arise from either two-dimensional
time-periodic fluid flows or three-dimensional magnetic fields, which are periodic in one direction.
The new method simply measures the stretching of a material curve
and the topological entropy is estimated as the exponential growth rate of the
line length.
By employing an adaptive approach we are able to study larger numbers of iterations
without loss of precision.
However, the direct methods should not be applied without restrictions.
For highly tangled fields (highly mixed flows) and for very long iterations the
computation time increases exponentially.
This is not the case for the \verb+braidlab+ package.
However, the \verb+braidlab+ performs significantly worse for the examples tested here
for which our method is faster.
In case of the $E1$ configuration with $\kappa = 1$ we find a speed up of a factor of ca.\ $450$
in favor of the direct adaptive method
on a 2.4 GHz quad-code Intel Xeon with 32 GB RAM.

The stretching of the initial curve, and hence the topological entropy,
is not homogeneous in the domain.
By measuring the lengthening of a distribution of circles, we show that the
finite time topological entropy can exhibit a complex spatial distribution.
For one of the maps used here, it shows a Cantor set like distribution, with
alternating zero and finite entropy.

This is confirmed by the mapping of a passive scalar for which we compute
its spatial distribution using a Fourier transform.
There we observe an exponential decrease in length scale for the passive scalar.
This was expected, since the mapping used exponentially increases any line length.

\acknowledgements
All the authors acknowledge financial support from the UK's
STFC (grant number ST/K000993).

\bibliographystyle{ieeetr}
\bibliography{references}

\begin{thebibliography}{10}

\bibitem{Adler-Konheim-1695-114-309-TrAmMathSoc}
R.~L. Adler, A.~G. Konheim, and M.~H. McAndrew, ``Topological entropy,'' {\em
  Trans. Amer. Math. Soc.}, vol.~114, pp.~309--319, 1965.

\bibitem{boyland2000}
P.~L. {Boyland}, H.~{Aref}, and M.~A. {Stremler}, ``{Topological fluid
  mechanics of stirring},'' {\em J.~Fluid Mech.}, vol.~403, pp.~277--304, 2000.

\bibitem{Budisic-Thiffeault-2015-25-8-Chaos}
M.~Budi\v{s}i{\' c} and J.-L. Thiffeault, ``Finite-time braiding exponents,''
  {\em Chaos}, vol.~25, no.~8, p.~087407, 2015.

\bibitem{Sattari-Chen-2016-26-3-Chaos}
S.~Sattari, Q.~Chen, and K.~A. Mitchell, ``Using heteroclinic orbits to
  quantify topological entropy in fluid flows,'' {\em Chaos}, vol.~26, no.~3,
  2016.

\bibitem{Petrisor-Misguich-2003-18-1085-ChaosSolFrac}
E.~Petrisor, J.~Misguich, and D.~Constantinescu, ``Reconnection in a global
  model of {P}oincar\'e map describing dynamics of magnetic field lines in a
  reversed shear tokamak,'' {\em Chaos, Solitons \& Fractals}, vol.~18, no.~5,
  pp.~1085 -- 1099, 2003.

\bibitem{Klapper}
I.~{Klapper} and L.~S. {Young}, ``{Rigorous bounds on the fast dynamo growth
  rate involving topological entropy},'' {\em Communications in Mathematical
  Physics}, vol.~173, pp.~623--646, Nov. 1995.

\bibitem{Childress}
S.~Childress, {\em Fast Dynamos}, pp.~313--327.
\newblock 1991.

\bibitem{Parker-1972-174-499-ApJ}
E.~N. {Parker}, ``{Topological Dissipation and the Small-Scale Fields in
  Turbulent Gases},'' {\em \apj}, vol.~174, p.~499, June 1972.

\bibitem{Berger-2001-55-3-LMathPhys}
M.~A. Berger, ``Topological invariants in braid theory,'' {\em Letters in
  Mathematical Physics}, vol.~55, no.~3, pp.~181--192, 2001.

\bibitem{pontin2011a}
D.~I. Pontin, A.~L. Wilmot-Smith, G.~Hornig, and K.~Galsgaard, ``{Dynamics of
  braided coronal loops. II. Cascade to multiple small-scale reconnection
  events},'' {\em Astron.~Astrophys.}, vol.~525, p.~A57, 2011.

\bibitem{Hansteen-Guerreiro-2015-811-106-ApJ}
V.~Hansteen, N.~Guerreiro, B.~D. Pontieu, and M.~Carlsson, ``Numerical
  simulations of coronal heating through footpoint braiding,'' {\em \apj},
  vol.~811, no.~2, p.~106, 2015.

\bibitem{Smiet-Candelaresi-2015-115-5-PRL}
C.~B. Smiet, S.~Candelaresi, A.~Thompson, J.~Swearngin, J.~W. Dalhuisen, and
  D.~Bouwmeester, ``Self-organizing knotted magnetic structures in plasma,''
  {\em Phys. Rev. Lett.}, vol.~115, p.~095001, Aug 2015.

\bibitem{Kroetz-Roberto-2012-54-4-PlasFus}
T.~Kroetz, M.~Roberto, I.~L. Caldas, R.~L. Viana, and P.~J. Morrison,
  ``Divertor map with freedom of geometry and safety factor profile,'' {\em
  Plasma Physics and Controlled Fusion}, vol.~54, no.~4, p.~045007, 2012.

\bibitem{Day-Frongillo-2008-7-4-SIAMAppDyn}
S.~Day, R.~Frongillo, and R.~Treviño, ``Algorithms for rigorous entropy bounds
  and symbolic dynamics,'' {\em SIAM Journal on Applied Dynamical Systems},
  vol.~7, no.~4, pp.~1477--1506, 2008.

\bibitem{Newhouse-Pignataro-1993-72-5-JStatPhys}
S.~Newhouse and T.~Pignataro, ``On the estimation of topological entropy,''
  {\em Journal of Statistical Physics}, vol.~72, no.~5-6, pp.~1331--1351, 1993.

\bibitem{Pesin-1977-32-55-RusMatSurv}
Y.~B. Pesin, ``Characteristic {L}yapunov exponents and smooth ergodic theory,''
  {\em Russian Mathematical Surveys}, vol.~32, no.~4, p.~55, 1977.

\bibitem{Young-2003-313-entropy}
L.-S. Young, ``Entropy in dynamical systems,'' in {\em Entropy}, pp.~313--327,
  Princeton University Press Princeton, NJ, 2003.

\bibitem{Ruelle-1978-9-83-BolSocBrasMat}
D.~Ruelle, ``An inequality for the entropy of differentiable maps,'' {\em
  Boletim da Sociedade Brasileira de Matem{\'a}tica - Bulletin/Brazilian
  Mathematical Society}, vol.~9, no.~1, pp.~83--87, 1978.

\bibitem{Dritschel-1989-10-77-CompPhysRep}
D.~G. Dritschel, ``Contour dynamics and contour surgery: Numerical algorithms
  for extended, high-resolution modelling of vortex dynamics in
  two-dimensional, inviscid, incompressible flows,'' {\em Computer Physics
  Reports}, vol.~10, no.~3, pp.~77 -- 146, 1989.

\bibitem{Mills-2009-35-2020-CompGeo}
P.~Mills, ``Isoline retrieval: An optimal sounding method for validation of
  advected contours,'' {\em Computers \& Geosciences}, vol.~35, no.~10,
  pp.~2020 -- 2031, 2009.

\bibitem{benettin1980}
G.~Benettin, L.~Galgani, A.~Giorgilli, and J.-M. Strelcyn, ``Lyapunov
  characteristic exponents for smooth dynamical systems and for {H}amiltonian
  systems; a method for computing all of them. ii - numerical application,''
  {\em Meccanica}, vol.~15, no.~1, pp.~21--30, 1980.

\bibitem{Thiffeault-2010-20-1-Chaos}
J.-L. Thiffeault, ``Braids of entangled particle trajectories,'' {\em Chaos},
  vol.~20, no.~1, p.~017516, 2010.

\bibitem{moussafir2006}
J.~O. Moussafir, ``{On computing the entropy of braids},'' {\em Functional
  Analysis and Other Mathematics}, vol.~1, pp.~37--46, 2006.

\bibitem{braidlab}
J.-L. Thiffeault and M.~Budi\v{s}i{\' c}, ``Braidlab: A software package for
  braids and loops,'' {\em arXiv:1410.0849}.

\bibitem{Henon-1976-69-50-CommMathPhys}
M.~H{\' e}non, ``A two-dimensional mapping with a strange attractor,'' {\em
  Comm. Math. Phys.}, vol.~50, no.~1, pp.~69--77, 1976.

\bibitem{pontin2016a}
D.~I. {Pontin}, S.~{Candelaresi}, A.~J.~B. {Russell}, and G.~{Hornig},
  ``{Braided magnetic fields: equilibria, relaxation and heating},'' {\em
  Plasma Phys.~Control.~Fusion}, vol.~58, no.~5, p.~054008, 2016.

\bibitem{Greene-1979-20-6-JMathPhys}
J.~M. Greene, ``A method for determining a stochastic transition,'' {\em
  Journal of Mathematical Physics}, vol.~20, no.~6, pp.~1183--1201, 1979.

\bibitem{Morrison-2000-7-6-PhysPlasm}
P.~J. Morrison, ``Magnetic field lines, {H}amiltonian dynamics, and nontwist
  systems,'' {\em Physics of Plasmas}, vol.~7, no.~6, pp.~2279--2289, 2000.

\bibitem{brunton2010}
S.~L. Brunton and C.~W. Rowley, ``Fast computation of finite-time {L}yapunov
  exponent fields for unsteady flows,'' {\em Chaos}, vol.~20, no.~1, 2010.

\bibitem{Pierrehumbert1991}
R.~T. Pierrehumbert, ``Large‐scale horizontal mixing in planetary
  atmospheres,'' {\em Physics of Fluids A}, vol.~3, no.~5, pp.~1250--1260,
  1991.

\bibitem{Szezech-Caldas-2012-86-8-PRE}
J.~D. Szezech, I.~L. Caldas, S.~R. Lopes, P.~J. Morrison, and R.~L. Viana,
  ``Effective transport barriers in nontwist systems,'' {\em Phys.\ Rev.\ E},
  vol.~86, p.~036206, 2012.

\bibitem{Szezech-Schelin-2013-377-452-PhysLetA}
J.~Szezech, A.~Schelin, I.~Caldas, S.~Lopes, P.~Morrison, and R.~Viana,
  ``Finite-time rotation number: A fast indicator for chaotic dynamical
  structures,'' {\em Phys.\ Lett.\ A}, vol.~377, no.~6, pp.~452 -- 456, 2013.

\bibitem{Cincotta-Simo-2000-147-205-AA}
{Cincotta, P. M.} and {Sim\'o, C.}, ``Simple tools to study global dynamics in
  non-axisymmetric galactic potentials – {I},'' {\em Astron.\ Astrophys.\
  Sup.}, vol.~147, no.~2, pp.~205--228, 2000.

\bibitem{Skokos-Bountis-2007-231-30-PhysD}
C.~Skokos, T.~Bountis, and C.~Antonopoulos, ``Geometrical properties of local
  dynamics in {H}amiltonian systems: {T}he {G}eneralized {A}lignment {I}ndex
  ({GALI}) method,'' {\em Physica D}, vol.~231, no.~1, pp.~30 -- 54, 2007.

\bibitem{Haller-Yuan-2000-147-352-PhysD}
G.~Haller and G.~Yuan, ``Lagrangian coherent structures and mixing in
  two-dimensional turbulence,'' {\em Physica D}, vol.~147, no.~3–4, pp.~352
  -- 370, 2000.

\bibitem{Lekien-Shadden-2007-48-6-JMatPhys}
F.~Lekien, S.~C. Shadden, and J.~E. Marsden, ``Lagrangian coherent structures
  in n-dimensional systems,'' {\em J.\ Math.\ Phys.}, vol.~48, no.~6,
  p.~065404, 2007.

\bibitem{Froyland2012}
G.~Froyland and K.~Padberg-Gehle, ``Finite-time entropy: A probabilistic
  approach for measuring nonlinear stretching,'' {\em Physica D}, vol.~241,
  no.~19, pp.~1612 -- 1628, 2012.

\bibitem{lesne2014}
A.~Lesne, ``Shannon entropy: a rigorous notion at the crossroads between
  probability, information theory, dynamical systems and statistical physics,''
  {\em Mathematical Structures in Computer Science}, vol.~24, no.~3, 2014.

\end{thebibliography}

\end{document}